\documentclass[twocolumn]{aastex61}
\usepackage{amsmath,amstext}
\usepackage[T1]{fontenc}
\usepackage{apjfonts} 
\usepackage[figure,figure*]{hypcap}
\usepackage{hyperref}

\shorttitle{GS2 1406}
\shortauthors{Larson, R.L. et al.}

\begin{document}

\title{Discovery of a \MakeLowercase{z} $=$ 7.452 High Equivalent Width Lyman-$\alpha$ Emitter from the {\it Hubble Space Telescope} Faint Infrared Grism Survey}

\author{Rebecca L. Larson}
\affiliation{The University of Texas at Austin, Department of Astronomy, Austin, TX, United States.}
\author{Steven L. Finkelstein}
\affiliation{The University of Texas at Austin, Department of Astronomy, Austin, TX, United States.}
\author{Norbert Pirzkal}
\affiliation{Space Telescope Science Institute, 3700 San Martin Dr., Baltimore MD 21218}
\author{Russell Ryan}
\affiliation{Space Telescope Science Institute, 3700 San Martin Dr., Baltimore MD 21218}
\author{Vithal Tilvi}
\affiliation{Arizona State University, Department of Physics and Astronomy, Tempe, AZ, United States.}
\author{Sangeeta Malhotra}
\affiliation{Arizona State University, Department of Physics and Astronomy, Tempe, AZ, United States.}
\affiliation{NASA Goddard Space Flight Center, Greenbelt, MD, United States}
\author{James Rhoads}
\affiliation{Arizona State University, Department of Physics and Astronomy, Tempe, AZ, United States.}
\affiliation{NASA Goddard Space Flight Center, Greenbelt, MD, United States}
\author{Keely Finkelstein}
\affiliation{The University of Texas at Austin, Department of Astronomy, Austin, TX, United States.}
\author{Intae Jung}
\affiliation{The University of Texas at Austin, Department of Astronomy, Austin, TX, United States.}
\author{Lise Christensen}
\affiliation{Dark Cosmology Centre, Niels Bohr Institute, University of Copenhagen, Juliane Maries Vej 30, DK-2100 Copenhagen, Denmark}
\author{Andrea Cimatti}
\affiliation{University of Bologna, Department of Physics and Astronomy (DIFA),
Via Gobetti 93/2, I-40129, Bologna,  Italy.}
\affiliation{INAF - Osservatorio Astrofisico di Arcetri, Largo E. Fermi 5
I-50125, Firenze, Italy.}
\author{Ignacio Ferreras}
\affiliation{Mullard Space Science Laboratory, University College London, Holmbury St Mary, Dorking, Surrey RH5 6NT, UK}
\author{Norman Grogin}
\affiliation{Space Telescope Science Institute, 3700 San Martin Dr., Baltimore MD 21218}
\author{Anton M. Koekemoer}
\affiliation{Space Telescope Science Institute, 3700 San Martin Dr., Baltimore MD 21218}
\author{Nimish Hathi}
\affiliation{Space Telescope Science Institute, 3700 San Martin Dr., Baltimore MD 21218}
\author{Robert O'Connell}
\affiliation{The University of Virginia, Charlottesville, VA, 22904-4325, USA}
\author{G\"oran \"Ostlin}
\affiliation{Stockholm University, Stockholm, SE-10691, Sweden}
\author{Anna Pasquali}
\affiliation{Astronomisches Rechen Institut, Zentrum fuer Astronomie, Universitaet 
Heidelberg, Moenchhofstrasse 12 - 14, D-69120 Heidelberg, Germany.}
\author{Barry Rothberg}
\affiliation{Large Binocular Telescope Observatory, 933 N. Cherry Ave Tucson AZ 85721}
\affiliation{George Mason University, Department of Physics \& Astronomy, MS 3F3, 4400 University Drive, Fairfax, VA 22030, USA}
\author{Rogier A. Windhorst}
\affiliation{Arizona State University, Department of Physics and Astronomy, Tempe, AZ, United States.}
\author{The FIGS Team}

\begin{abstract}
We present the results of an unbiased search for Ly$\alpha$ emission from continuum-selected 6 $< z <$ 8 galaxies. Our dataset consists of 160 orbits of G102 slitless grism spectroscopy obtained with the \textit{Hubble Space Telescope} (\textit{HST}) Wide Field Camera 3 as part of the Faint Infrared Grism Survey (FIGS; PI: Malhotra), which obtains deep slitless spectra of all sources in four fields, and was designed to minimize contamination in observations of previously-identified high-redshift galaxy candidates. The FIGS data can potentially spectroscopically confirm the redshifts of galaxies, and as Ly$\alpha$ emission is resonantly scattered by neutral gas, FIGS can also constrain the ionization state of the intergalactic medium (IGM) during the epoch of reionization. These data have sufficient depth to detect Ly$\alpha$ emission in this epoch, as \citet{Tilvi2016} have published the FIGS detection of previously known \citep{Finkelstein2013} Ly$\alpha$ emission at $z=$7.51.  The FIGS data use five separate roll-angles of \textit{HST} to mitigate the contamination by nearby galaxies. We created a method that accounts for and removes the contamination from surrounding galaxies, and also removes any dispersed continuum light from each individual spectrum \citep{Pirzkal2017}. We searched for significant ($>4\sigma$) emission lines using two different automated detection methods, free of any visual inspection biases. Applying these methods on photometrically-selected high-redshift candidates between $ 6 < z < 8$ we find two emission lines, one previously published by \citet{Tilvi2016}, and a new line at 1.028 $\mu$m.  We identify this lines as Ly$\alpha$ at $z = 7.452\pm0.003$. This newly spectroscopically confirmed galaxy has the highest Ly$\alpha$ rest-frame equivalent width (EW$_{Ly\alpha}$) yet published at $z >$ 7 (140.3$\pm$19.0\AA).
\end{abstract}

%\keywords{keyword1 --- keyword2 --- keyword3}

\section{Introduction}
\label{sec:intro}

While thousands of candidate galaxies have been discovered in the epoch of reionization at $z >$ 6 using photometric measurements (e.g., \citealt{Finkelstein2015, Bouwens2015, McLeod2015, Bowler2015, Ono2017}), spectroscopic information is very limited. Accurate distance measures are required to improve our understanding of the evolution of galaxies, as any uncertainty in the redshift propagates through to uncertainties in key physical properties such as the luminosity, stellar mass, and star-formation rate.  These physical quantities are used to constrain theoretical models of galaxy formation and evolution, thus it is necessary to measure spectroscopic redshifts for a representative sample of photometrically selected galaxies, both to measure the contaminant fraction, and to calibrate the photometric redshift uncertainties.

At $z > 3$, Ly$\alpha$ emission is the dominant observed spectral feature used to search for galaxies \citep{Rhoads2000,Kudritzki2000} because it is the emission line most accessible from ground-based observations (e.g., \citealt{Finkelstein2016, Stark2016}, and references therein). 
Selecting galaxies by their Ly$\alpha$ emission through narrowband surveys (e.g., \citealt{Hu1998, Rhoads2000, Kudritzki2000, Steidel2000, Ouchi2003}) and direct spectroscopic searches (e.g., \citealt{Malhotra2005, Pirzkal2007, Rhoads2013}) identifies populations that at most evolve weakly from $z\approx 3$ to $z\approx 6$, whether in Ly$\alpha$ luminosity \citep{Dawson2007} or in UV size and surface brightness \citep{Malhotra2012}. The line strengths of Ly$\alpha$-selected samples are large \citep{Malhotra2002} and detectably evolve from smaller equivalent widths at $z\approx 3$ to larger ones at $z\approx 6$ \citep{Zheng2014}.  Similarly, the fraction of continuum-selected galaxies (e.g., Lyman break galaxies) which have detectable Ly$\alpha$ emission via follow-up spectroscopy rises from $\sim 30\%$ at $z = 3$ to 60-80\% at $z = 6$ (\citealt{Shapley2003, Stark2010, Stark2011}, though see also \citealt{Caruana2018}). This implies that Ly$\alpha$ should be both a powerful and efficient means of measuring the redshifts to galaxies at $z \sim 7$ and beyond. 

However, the observations at $z>7$ tell a more complicated story. The number of Lyman break selected galaxies spectroscopically confirmed via Ly$\alpha$ at $z > 7$ is about a dozen (e.g., \citealt{Fontana2010, Vanzella2011, Shibuya2012, Schenker2012, Pentericci2014, Oesch2015, Zitrin2015}), with only five confirmed Ly$\alpha$ lines at $z > 7.5$ \citep{Finkelstein2013, Oesch2015, Zitrin2015, Song2016, LaPorte2017}. This is significantly fewer than expected based on the numbers of LBG candidates observed. Narrowband surveys continue to successfully identify Ly$\alpha$ lines at $z\approx 7.0$ \citep{Iye2006,Zheng2017}, $z\approx 7.3$ \citep{Shibuya2012, Konno2014}, and $z\approx 7.7$ (Tilvi et al 2018, in prep), but here too the numbers are generally lower than expected based on observations of the $z < 6$ universe.

A decrease in observable Ly$\alpha$ lines was anticipated as a likely consequence of neutral intergalactic gas prior to reionization \citep{Malhotra2004}.  Reionization history remains substantially unknown, but Ly$\alpha$ emission serves as a powerful probe, because neutral fractions over $\sim 30\%$ will scatter enough Ly$\alpha$ photons out of the line-of-sight to render detections difficult. The attenuation of Ly$\alpha$ lines was first used as a reionization test by \citet{Malhotra2004}, who found that narrowband Ly$\alpha$ observations then available were inconsistent with a fully neutral IGM at $z\approx 6.5$. Corresponding efforts using Ly$\alpha$ follow-up of $z>7$ Lyman break selected candidates were first published in 2011 \citep{Pentericci2011, Ono2012, Schenker2012}, and showed a significant deficit in Ly$\alpha$ lines.  While it is possible that the lack of spectroscopic detections could indicate a flaw in the selection process, this is unlikely as the method for selection (via the Lyman break) is identical to that used at lower redshifts, where the contamination rate has been determined to be quite low (e.g., \citealt{Pentericci2011}).  Rather, this change in Ly$\alpha$ detectability is likely related to residual neutral IGM, though evolution in galaxy gas properties could also play a role \citep{Finkelstein2012}. Currently, we know that the midpoint of reionization occurred around $z=8.8$ \citep{Planck2014}, and is largely complete by $z \sim 6$ (e.g., \citealt{Malhotra2004, Fan2006, Becker2015}); but the detailed history of reionization remains substantially unknown.  Searching for Ly$\alpha$ emission in Lyman break galaxies at $7 < z < 9$ is a powerful way to move forward, as even non-detections of Ly$\alpha$ can be constraining.

The apparent paucity of Ly$\alpha$ detections at $z > 6.5$ has led to a number of analyses on the neutral fraction, with some studies finding an IGM neutral fraction as high as 50-70\% at $z \sim 7$ (from $\sim$ fully ionized at $z \sim 6$; \citealt{Pentericci2011, Treu2013, Tilvi2014, Mason2017}). However, there are a variety of effects which can reduce the ability of observations to make an impact. The most pressing is that at $z \sim$ 6 -- 8, the photometric redshift probability distribution functions straddle the boundary between optical and near-infrared cameras, making it difficult with one instrument to probe the full wavelength range where a line may be found. This is compounded by the increasing sky brightness, and bright telluric emission and absorption features at these and longer wavelengths, further reducing the discovery space. Both of these effects can be mitigated with space-based slitless grism spectroscopy with the Hubble Space Telescope (HST). The HST Wide Field Camera 3 (WFC3) G102 grism covers the range 0.8$\mu m$ -- 1.15$\mu m$ at a spectral resolution of R$\sim$210, fully covering Ly$\alpha$ emission at $5.6 < z < 8.5$, throughout the epoch in question, all free of telluric emission lines (though not scattered earthshine). Grism spectra have previously been used to successfully detect both Lyman break galaxies \citep{Malhotra2005,Rhoads2009,Oesch2015} and emission line galaxies \citep{Malhotra2005, Rhoads2009, Rhoads2013, Pirzkal2007, Schmidt2017, Bagley2017} at high redshifts.

In this paper we report on a search for Ly$\alpha$ emission from galaxies in this epoch with data from the deepest {\it HST} grism survey yet, the Faint Infrared Grism Survey (FIGS; PI: Malhotra; \citealt{Pirzkal2017}).  We describe FIGS in \S 2, and outline our method for data reduction and emission line discovery in \S 3.  We summarize our results in \S 4, and discuss the implications in \S 5, and our conclusions in \S 6.  All magnitudes are given in the AB magnitude system \citep{Oke1983} and we assume H$_\mathrm{0}$ = 67.3 km s$^{-1}$ Mpc$^{-1}$, $\Omega_m = 0.315$ and $\Omega_{\Lambda} = 0.685$ \citep{Planck2015}.

\section{Data}
\label{sec:data}

%We use extremely deep \textit{Hubble Space Telescope} (\textit{HST}) grism data from  to detect Ly$\alpha$ emission lines from galaxies. 
FIGS is currently the most sensitive \textit{HST} G102 grism survey, and targets the Cosmic Assembly Near-infrared Deep Extragalactic Legacy Survey (CANDELS: \citealt{Grogin2011, Koekemoer2011}) Great Observatories Origins Deep Survey (GOODS: \citealt{Giavalisco2004}) fields. The \textit{HST} WFC3 grism  is used for obtaining slitless spectroscopy of an entire 123"$\times$136" field of view. This gives us spectra for $\sim$6,000 galaxies across four fields, complete to J$\sim$26.5 magnitude. The unavailability of a slit leads to contamination of nearby sources as the light is spread out along the dispersion axis, and in order to remove this effect, the same field is observed at different roll angles, changing the axis of dispersion.  This changes the amount of, if not completely avoiding, contamination of light from nearby sources that might fall into the dispersion pattern of a given object. The FIGS survey consists of 4 \textit{HST} pointings, each with 40 orbits spread over 5 different position angles in an effort to reduce the overall contamination effects from spatially nearby galaxies and foreground stars. The full description of this survey is available in \citet{Pirzkal2017}. This data set has already proven to be successful as \citet{Tilvi2016} detail the FIGS detection of a previously-known Ly$\alpha$ emitter at $z =$ 7.51 (\citealt{Finkelstein2013}).

\section{Method}

\subsection{Reduction from Raw Data to 2D Spectra}
\label{sec:2d}

\begin{figure*}[ht]
{\includegraphics[]{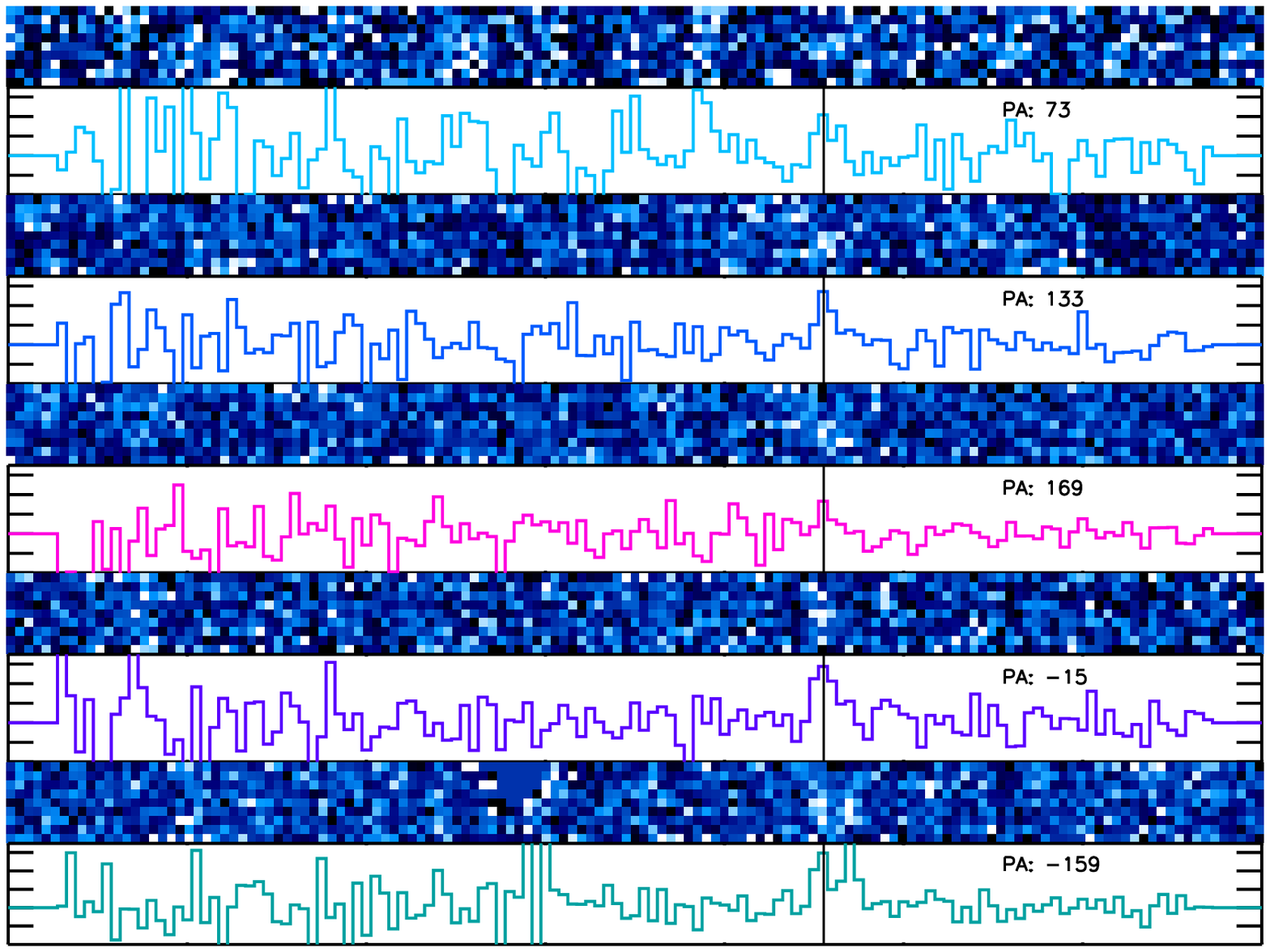}
\includegraphics[]{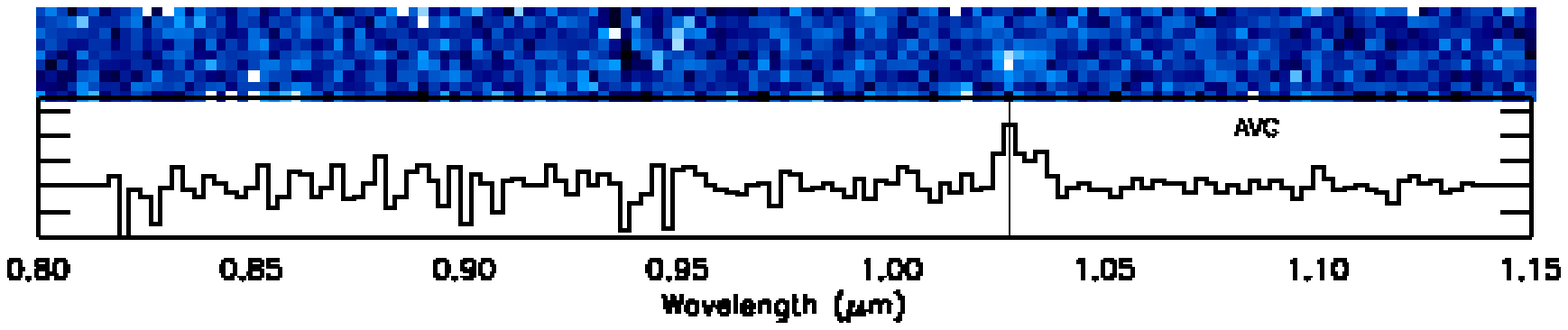}}
{\caption{2D and 1D, contamination-subtracted, spectra of GS2\_1406 in the FIGS data set. Top 5 plots are 2D/1D spectra from individual position angles, and bottom is a weighted combination of all 5 position angles, all after the reductions from \citet{Pirzkal2017}. This galaxy exhibits a strong emission line near 1.03$\mu$m which appears as a bright spot in the right half of each 2D spectrum, and which is marked by a vertical line in each 1D spectrum.}
\label{fig:2D1D}}
\end{figure*}

The method of reduction from raw data to two-dimensional (2D) spectra for each galaxy is explained in \citet{Pirzkal2017}, and follows loosely the method for the Advanced Camera for Surveys (ACS) programs Grism-ACS Program for Extragalactic Science (GRAPES: \citealt{Pirzkal2004}) and Probing Evolution and Reionization Spectroscopically (PEARS: \citealt{Pirzkal2009}), but is discussed here briefly for completeness. These reductions relied on first being able to simulate the data, and thus Simulation Based Extractions (SBEs) were performed using the full-depth \textit{HST} ACS and WFC3 mosaics in this field. This imaging is predominately from the CANDELS \citep{Grogin2011, Koekemoer2011} and GOODS \citep{Giavalisco2004} surveys and information on how these mosaics were created can be found in \citet{Koekemoer2011}. Hot and cold catalogs were created from these mosaics using a custom version of Source Extractor \citep{Bertin1996} with elliptical Kron \citep{Kron1980} apertures similar to the catalogs from \citet{Finkelstein2010, Finkelstein2012, Finkelstein2015}. Cold catalogs were created first to find the brighter extended sources, while more aggressive hot catalogs were created for fainter, smaller sources. A final catalog was taken by adding the sources from the hot catalog to the cold catalog as long as they did not overlap a cold-catalog object's segmentation map.

The publicly available WFC3 G102 grism calibration file \citep{Pirzkal2016, Pirzkal2017a} was used to simulate every single FIGS grism observation with special software that dispersed every object-pixel from the mosaic into the reference frame of the FIGS observation. This allows for the calculation of the dispersion solution and creation of data cubes which were used to generate simulated dispersed images for all objects. These simulated images and data cubes were used to determine which pixels in the real observations were needed to produce 2D spectra for each object. In this way it was possible to determine the dispersed flux in each pixel from nearby sources in the field and get a measure of contamination for each object. 

The combination of all this information was used to create a 2D, wavelength-rectified image for each object, by binning the data for each galaxy in 25{\AA} bins based on the properties of the G102 filter. The error values are computed as the RMS of the multiple ($\sim$32) measurements used in creating the 2D spectrum. Having five roll angle observations and subsequent simulations of the dispersion solutions for all objects in the field allowed for creation of five 2D, contamination-subtracted spectra for every galaxy. This process also created 2D models of where each galaxy was expected to be spatially (z-direction) dispersed by combining the Source Extractor footprint with the broadband photometry. 2D weighted maps were created for each galaxy using these models. We do not combine the five 2D maps because the object profile in the dispersion direction defines the resolution of each spectrum and these are different across position angles. They also each have different background residuals and contamination effects so we analyze them separately, or after these effects are corrected for as in the following reduction steps. An example of the five 2D spectra for the most robust detection of a $z >$ 7  galaxy (FIGS ID: GS2\_1406) are shown in Figure \ref{fig:2D1D}. For more detail on this process see \citet{Pirzkal2017}.

\subsection{Reduction from 2D to 1D Spectra}
\label{sec:1d}
The extraction from 2D spectra to one-dimensional (1D) spectra is done using the optimal extraction technique from \citet{Horne1986}. This method applies non-uniform weights to pixels in the spatial direction based on the photometric shape of the object to achieve better spectrophotometric accuracy. In the reduction process done by \citet{Pirzkal2017} several 2D products are created for each galaxy for each position angle of the telescope: the contamination-subtracted spectrum (S), a spatial profile of the object in the 2D spectrum (W), and an error (E). We extract the 2D spectrum into a 1D one by spatially (z-direction) summing per wavelength pixel using a simplified version of the optimal extraction equation from \citet{Horne1986}: $$f_{opt} = \frac{\Sigma_{z} (S W) / E^{2}}{\Sigma_{z} W^{2}/E^{2}}$$ which gives us an inverse-variance weighted optimal flux ($f_{opt}$) value at each wavelength pixel. We then take this 1D spectrum, apply the sensitivity curve of the G102 grism, and use this final spectrum for all following work (see Figure \ref{fig:2D1D} for 1D optimally-extracted spectra for the $z =$ 7.452 object). This process is also illustrated in \citet{Pirzkal2017}.

%\begin{figure}[ht]
%{\includegraphics[width=8.5cm]{GS2_1406_ogspeccolumn.eps}}
%{\caption{1D, contamination-subtracted, spectra of GS2\_1406 in the FIGS data set extracted from the 2D spectra in Figure \ref{fig:2D}. The top 5 plots are 2D spectra from individual position angles, and bottom is a weighted combination of all 5 position angles. This galaxy shows a strong emission line near 1.02$\mu$m (vertical line in each spectrum shows location of emission line).}
%\label{fig:1D}}
%\end{figure}

\subsection{Line-Fitting Routine}
\label{sec:routine}
As we are looking for Ly$\alpha$ emission from high-redshift galaxies, we focused on a subset of galaxies in the FIGS data that were previously classified, with CANDELS photometry, to be at $z >$ 5.5 \citep{Finkelstein2015}. This sample consists of 154 galaxies in our four fields, 24 of which are brighter than $J$=26.5, and could potentially be detected in our data. We restrict our analysis to the section of each spectra between 8,500 - 11,200{\AA}, as the sensitivity curve of the G102 instrument drops off significantly outside this range, substantially increasing the noise.

Rather than relying on uncertain and arbitrary visual inspection of 2D spectra to identify plausible emission lines, we utilize a Monte Carlo Markov Chain (MCMC) routine (Ryan et. al, in prep) to search for significant emission lines in the 1D spectra. While we know Ly$\alpha$ has an asymmetric profile, at this spectral resolution (R$\sim$210) we do not expect to resolve this asymmetry and a gaussian function is an appropriate fit to this data. As such, our fitting routine fits a gaussian + constant function that takes in four parameters: the spectroscopic continuum level constant, central wavelength, full-width half-max (FWHM), and integrated line flux. We use an IDL implementation of the affine-invariant sampler \citep{Goodman2010} to sample the {\it posterior}, which is similar to the \texttt{emcee} package \citep{Foreman-Mackey2013}. We run the MCMC code with 500,000 iterations and 100 walkers at each pixel (significantly past the convergence point), stepping through wavelength space. This allows for a mostly unbiased search as we are fitting a Gaussian centered at every wavelength pixel across the spectrum, instead of giving an expected location for our emission line based on the photometric redshift information from \citet{Finkelstein2015}. A comparison to photometric redshifts for detected lines is discussed in Section \ref{sec:photometry}. Fitting parameters for the initial run are shown in Table \ref{tab:param}.

%We use the term ``blind" because we do not assign any prior information about the galaxy redshift or continuum flux as measured from the photometry.

\begin{table}
\begin{tabular}{ |p{3.5cm}p{4cm}| }
 \hline
 \multicolumn{2}{|c|}{Fitting Parameters} \\
 \hline
 "Continuum" Constant &  $-1\times10^{-18}$ {\textless} C {\textless} $1\times10^{-18}$ erg s$^{-1}$ cm$^{-2}$ \AA$^{-1}$ \\
 Peak Wavelength      &  $\lambda_{\rm pixel} \pm 12$\AA  \\
 Gaussian FWHM        &  25{\AA} {\textless} FWHM {\textless} 68 {\AA}  \\
 Line Flux            &  $10^{-20}$ {\textless} Flux {\textless} $10^{-15}$    erg s$^{-1}$ cm$^{-2}$\\
 \hline
\end{tabular}
\caption{Fitting parameters for the MCMC chain that fits a constant + gaussian function at each wavelength pixel. These parameters are used in the first fit to each position angle spectrum. Once residual contamination is removed (\S 3.3.1) the continuum constant is fixed to 0.}\label{tab:param}
\end{table}

\subsubsection{Removal of Residual Contamination}
\label{sec:residual}
While the data reduction process takes into account much of the contamination and residual emission from nearby sources, there is often an overall zeroth order continuum shape to each spectrum.  This is likely due to residual contamination that is missed during these reduction steps. Galaxies at these redshifts should have very faint continuum emission, and by removing any residual continuum shape to the spectrum we are not significantly affecting emission line results, but are accounting for imperfect noise and contamination corrections done in earlier steps. Searches for real continuum breaks in these data are discussed in \citet{Tilvi2016}.

The first step of this process is to use our MCMC routine to fit a gaussian function + a constant centered at each pixel of the wavelength array. At this step we let the constant vary between $\pm1\times10^{-18}$ erg s$^{-1}$ cm$^{-2}$ \AA$^{-1}$, which are much larger values than the 1$\sigma$ noise level and are much higher than the typical continuum values for our high-redshift sources. We also restrict the peak wavelength to be the wavelength at that pixel $\pm12$\AA, such that we fit a gaussian within each pixel. We limit the FWHM to between 25{\AA}, which is the instrumental resolution, and the FWHM which would correspond to 2000 km s$^{-1}$ ($\sim 68$\AA) as calculated by $\rm FWHM_{max} = 2000\ km\ s^{-1}\ \frac{\lambda_{peak}}{c} $ (where c is the speed of light). We force the line flux value to be greater than $10^{-20}$ and less than $10^{-15}$ erg s$^{-1}$ cm$^{-2}$ which does not put strong restrictions on the MCMC chain, but keeps the chain from spending time in unlikely regions of parameter space. These parameters are listed in Table \ref{tab:param}. 

\begin{figure}[ht!]
{\includegraphics[width=.5\textwidth]{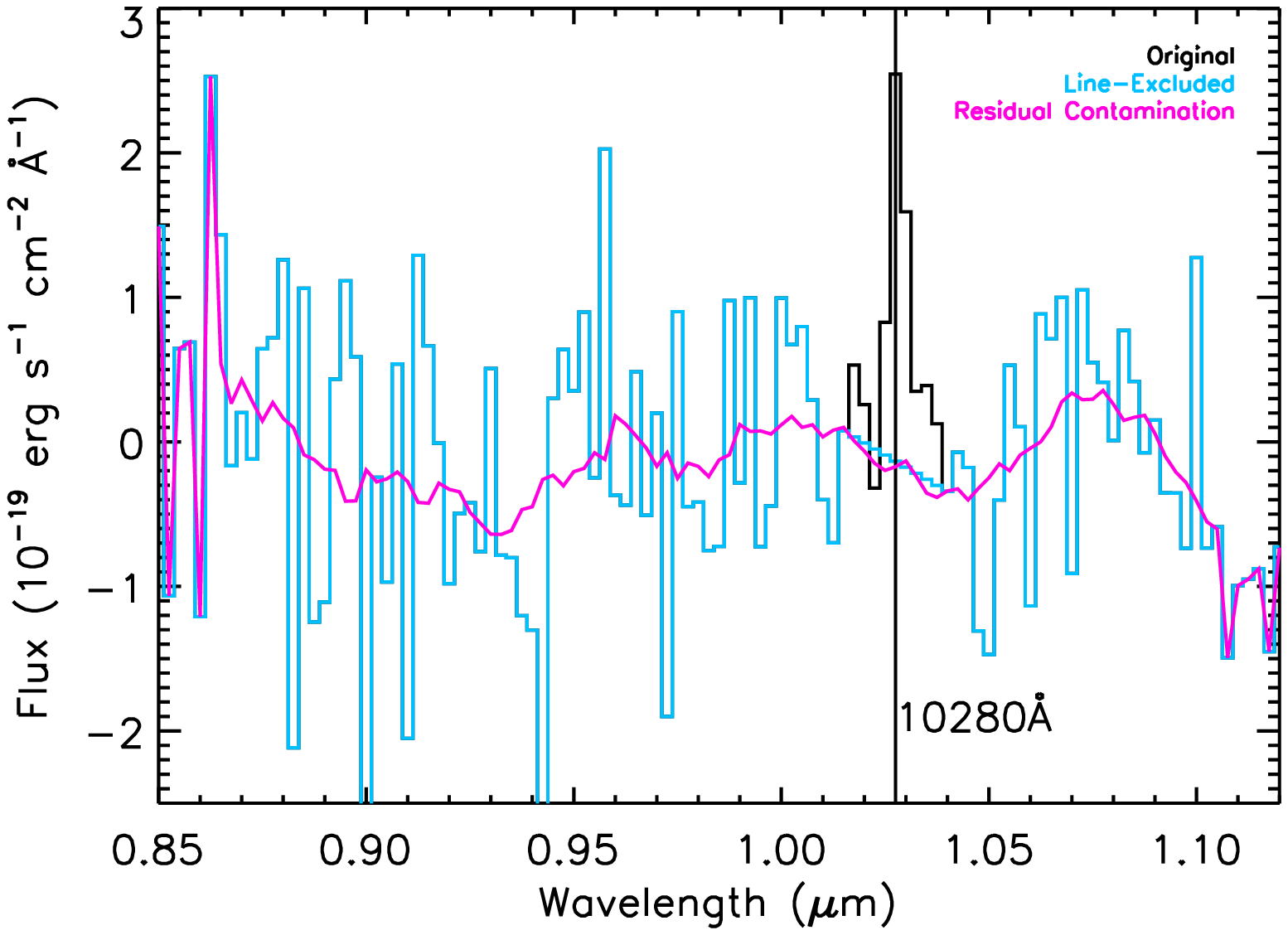}}
{\includegraphics[width=.5\textwidth]{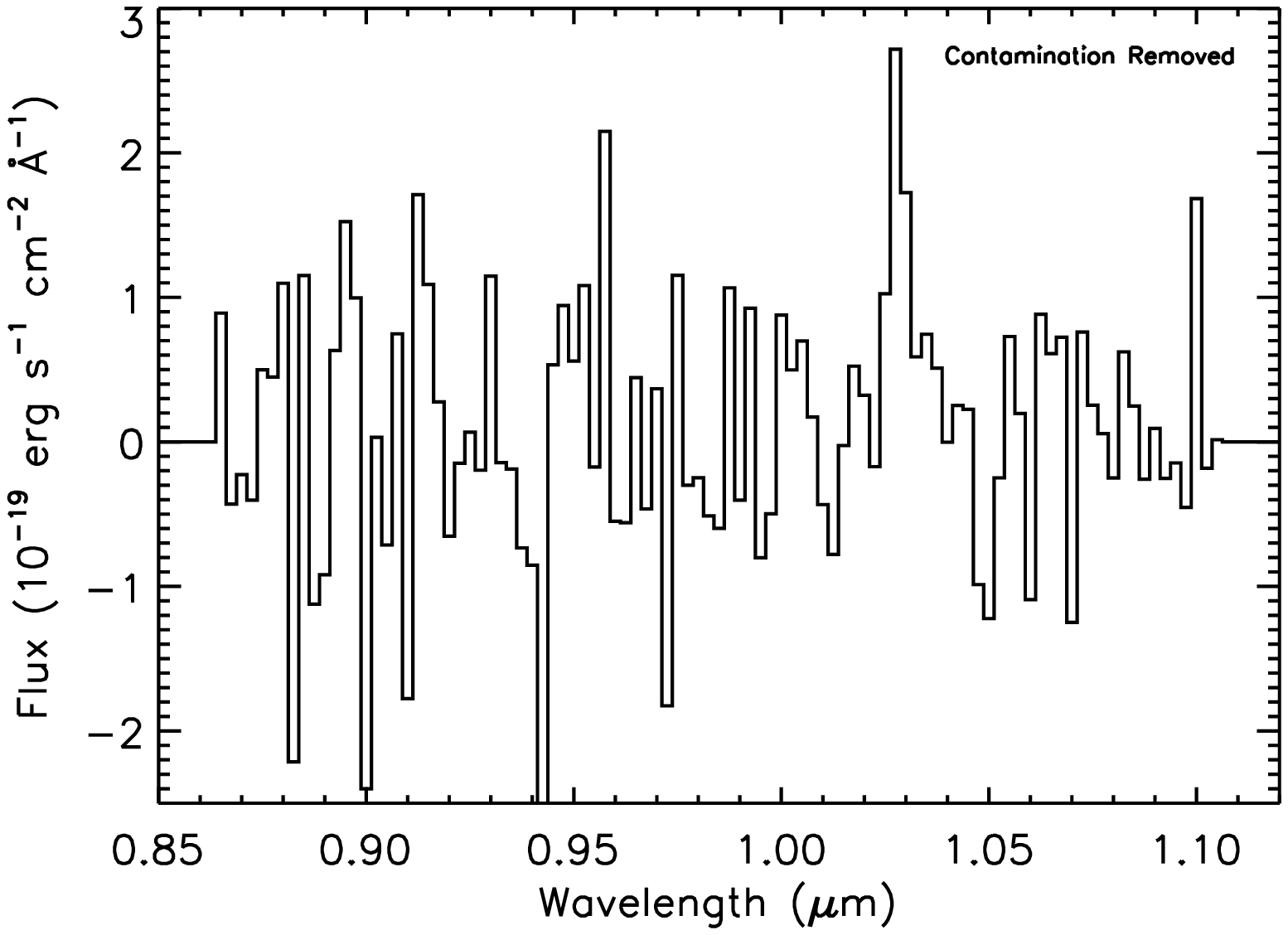}}
{\caption{{\bf Top:} Example of a residual contamination ("continuum") fit to one position angle spectrum. The blue line shows the original spectrum, while the black portion shows the region 3 pixels on either side of the peak of a line identified in the first pass (indicated by the vertical line). The pink line is our residual, which is calculated by 12-pixel boxcar smoothing of the blue line which interpolates across the potential emission line. 
{\bf Bottom:} The final "flattened" spectrum after this residual is subtracted.}
\label{fig:smooth}}
\end{figure}

To measure our line fluxes we use the median value of the last 100,000 steps of our MCMC chain, well after it has sufficiently converged. We use the ``robust sigma'' calculation to measure our error: using the median absolute deviation as the initial estimate, then weighting points using Tukey's Biweight (equation 9 from \citealt{Beers1990}). We calculate the signal-to-noise (SNR) of the emission line as the median line flux divided by the line flux error. We count the fit as a potential emission line if it has a SNR $>$ 4 and also has the lowest $\chi^2$ of the surrounding two pixels ($\pm25$\AA) on each side (accounting for a single line being detected in multiple pixels).

To remove this residual contamination we mask out any detected emission lines from our first pass (those with SNR $>$ 4) and a region around them ($\pm$ 3 pixels on either side) and then interpolate over these regions. Here we use a larger region than the expected FWHM of the emission line to insure we are not smoothing out the wings of the line profile. We then fit a boxcar smoothing function with a width of 12 pixels to the entire spectrum to average over the noise and identify a smooth residual component (see Figure \ref{fig:smooth}). It is possible that this results in a slight over or under subtracting of the residual contamination but this effect is minimal in the search for an emission line. Once we have measured this residual we subtract it from the original spectrum to produce our final spectrum (See Figure \ref{fig:smooth}, Bottom).

\begin{figure*}[ht!]
{\includegraphics[width=0.5\textwidth]{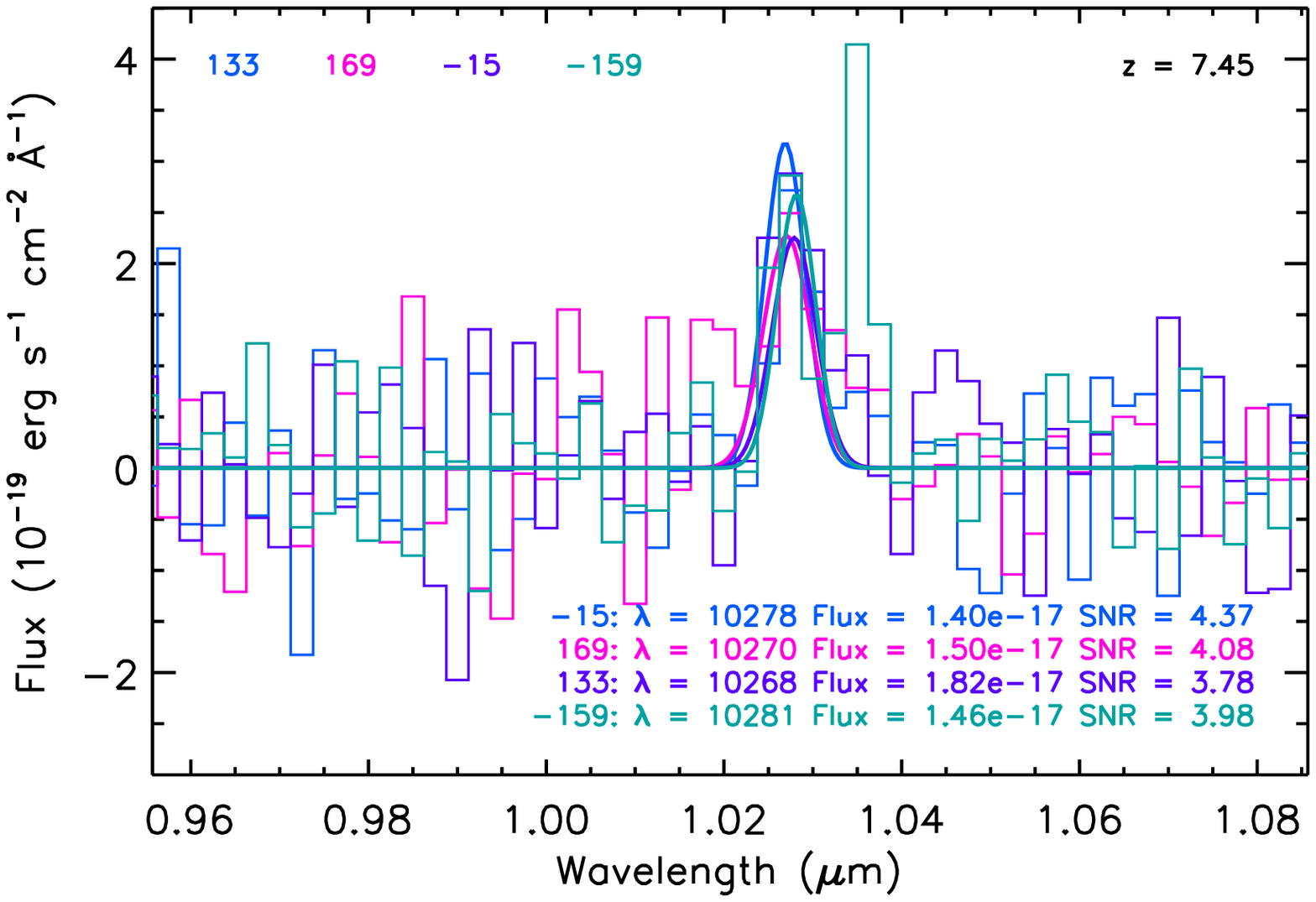}
\includegraphics[width=0.5\textwidth]{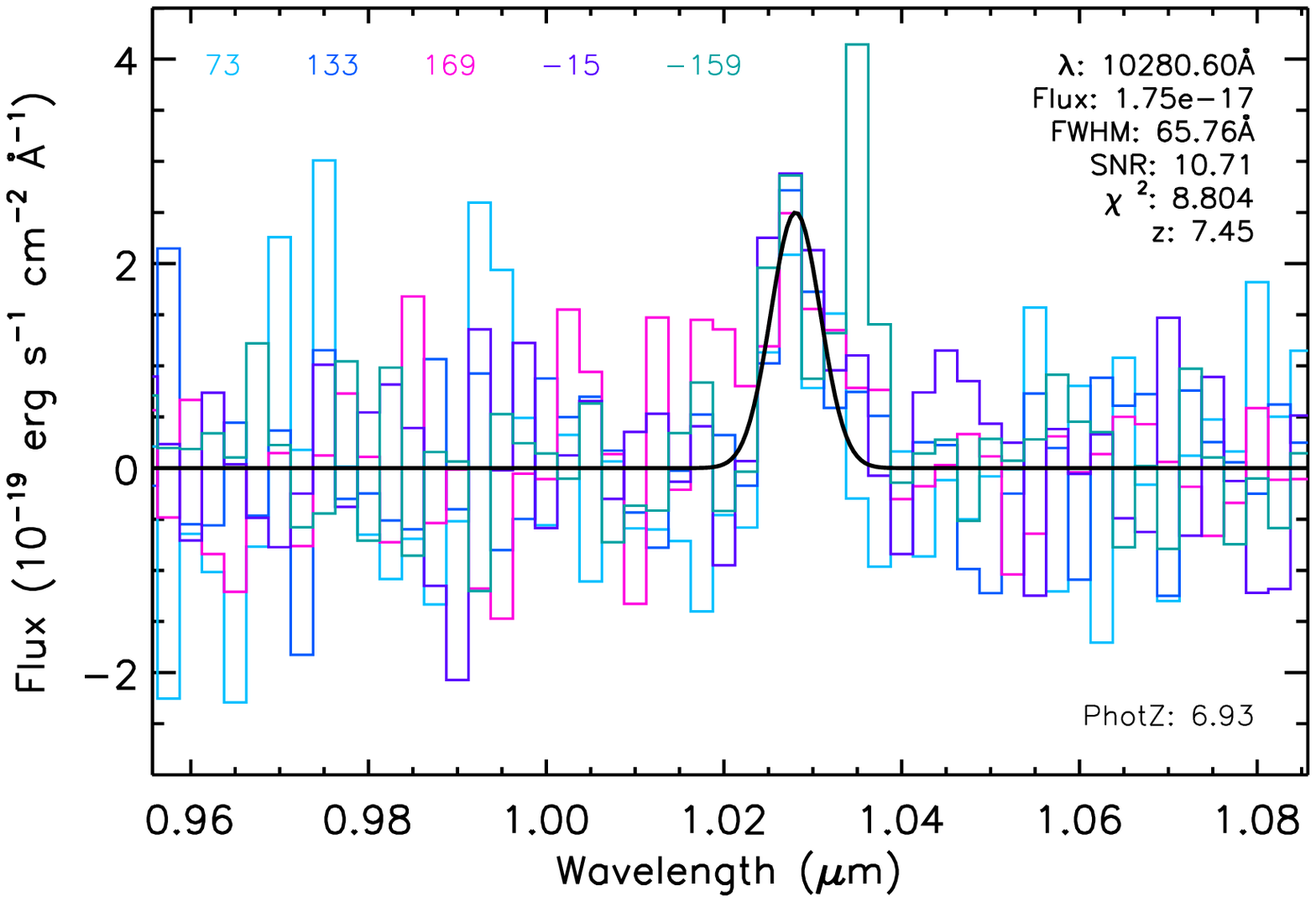}
}
{\caption{\textbf{Method 1 (Left):} Example of an emission line found in 4 position angle spectra (colored) individually for GS2\_1406 in the FIGS data. Individual PA fit results are printed as well as their corresponding SNR measurements. Note that our SNR cut for this method is $>4\sigma$ but we are showing all detections above $3\sigma$ in this plot to further illustrate the significance of this emission line. \textbf{Method 2 (Right):} Example of an emission line fit (black) to all five position angle spectra (colored) simultaneously for GS2\_1406. This emission line has a flux of $(1.75\pm0.16) \times 10^{-17}$ erg s$^{-1}$ cm$^{-2}$ and gives a redshift for the galaxy of $z = 7.452\pm0.003$. We use this measurement as the reported line flux information as it uses all the available spectra and more accurately accounts for potential noise amplification from one PA. }\label{fig:Methods}}
\end{figure*}

\subsubsection{Line Detection Steps}
We then search for emission lines in these fully reduced 1D spectra using two different and independent methods. First we look for matching lines in more than one position angle for the same galaxy. Second we simultaneously fit all five position angle spectra and use a combined $\chi^2$ estimate to find emission lines. Each method is described in the following sections but they both follow the same general steps. We use our MCMC routine to fit a gaussian at each wavelength pixel, using the same restrictions as in Table \ref{tab:param} except we now fix the constant to 0 for both as we expect there to be no remaining continuum emission after our steps in \S 3.3.1.

\subsubsection{Method 1: Matching Lines from Individual Position Angles}
\label{sec:method1}
In this method we run our line-finding code on each of the five spectra separately, searching for $> 4\sigma$ detections in individual position angles. In order for an emission line to be selected as a potential real emission line in this method, an object must have a line detected at $> 4\sigma$ significance in two or more position angles at the same wavelength $\pm$ 2 pixels ($\pm$ 50\AA). Finding a significant line in only one position angle could just be indicative of a noise spike or neighboring contamination, and setting the detection threshold at $4\sigma$ removes the detection of potentially correlated noise as the lines span several pixels. If the emission line is real, the rotation of the telescope will not affect the wavelength at which the emission line is found and therefore, searching for lines at matching wavelengths in more than one position angle provides further evidence of real detections. Here we assume that the emission line source is not offset from the assumed center of the object. This method finds 2 candidate emission line galaxies in all four of the FIGS fields. An example of a successful fit to four position angles of to GS2\_1406 can be found in the left panel of Figure \ref{fig:Methods}.

\subsubsection{Method 2: Fit to all Five Position Angles Simultaneously}
\label{sec:method3}
For this method we fit all five PAs simultaneously using the same fitting parameters as before, except now we are using the combined $\chi^2$ value of the same Gaussian fit to all five PAs as the goodness-of-fit statistic. Real lines might not be detected in all PAs but these PAs will have larger uncertainties and will thus be down-weighted in this method. An example of a fit to GS2\_1406 using this method is shown in the right panel of Figure \ref{fig:Methods}. This method finds 5 emission lines. This method is also the one we use for our final fit values for significant emission lines, as it includes all the available spectra and more accurately accounts for potential noise amplification from one PA.

\subsection{Method Validation}
In an effort to rule out the possibility that any of our detections were spurious we used both of these methods to fit a sample of spectra from 47 objects from the FIGS dataset in the GS1 field that are highly unlikely to have real emission lines, as these objects are extremely faint (m$\sim$29). Using both methods, with the same fitting criteria as above, we recovered \emph{no emission lines} and therefore conclude that the likely contamination rate of spurious noise being misidentified as a significant emission line in our sample is negligible. It is likely that some of our individual detected emission lines are in fact noise, but by invoking the criteria that they are found at the same wavelength in multiple PAs (Method 1) or that they are found in a simultaneous fit to all five PAs (Method 2) we are not including them in our results. The EM2D method (\citealt{Pirzkal2013} and Pirzkal et al. 2017, in prep) also uses a combination of two methods to identify emission line galaxies using the 2D spectra for this reason. 

We also tested both methods on low-redshift lines to determine the likelihood that a significant emission line exists in our data and we do not recover it. To do this we used a sample of known emission lines in the FIGS dataset from lower-redshift galaxies but with roughly the same fluxes as we expect Ly$\alpha$ to have in our high-redshift sample. These emission lines are identified as either H$\alpha$ or [O\,{\sc iii}] and are discussed in an upcoming paper by Pirzkal et. al. (2017, in prep). Of the 8 objects in this sample we recover a significant emission line using both methods in 7 of them. The 8th object has a brighter emission line than we included in our parameter space ($> 3.5 \times 10^{-16}$) and as such our method does not accurately fit this data. This one emission line is $\sim 30$ times brighter than the brightest line we find, and would expect to find, in our high-redshift sample and we thus exclude this from our test measures and determine that our code is accurately recovering significant emission lines in our dataset.

\section{Emission Line Results}
\label{sec:results}
We find 5 emission line galaxies in at least one method, and 2 galaxies in both methods. One of these galaxies detected by both methods is the known Ly$\alpha$ emission line at $z =$ 7.51 from \citet{Finkelstein2013} and \citet{Tilvi2016} (FIGS ID: GN1\_1292) which is found in our two methods at $>5\sigma$ significance: it is found in the two PAs as reported by \citet{Tilvi2016} and it is also found by fitting all five PAs simultaneously. Our measured line flux for this line is $(1.10\pm0.17)\times 10^{-17}$ erg s$^{-1}$ cm$^{-2}$ which is consistent with the measured value from \citet{Tilvi2016}, $(1.06\pm0.12)\times 10^{-17}$ erg s$^{-1}$ cm$^{-2}$, using this same dataset. As this line was originally identified as Ly$\alpha$ from ground-based Keck MOSFIRE spectra by \citet{Finkelstein2013}, this in part, validates our line identification procedure. 

For the remainder of this paper we focus on the second emission line selected via both methods, which has not been previously published.  This line is found in FIGS ID: GS2\_1406 (ID z7\_PAR2\_2909 in \citealt{Finkelstein2015}), at a position of $\alpha=$53.288090, $\delta=-$27.865408. This galaxy has a detected emission line at $10280.60\pm3.94${\AA} with a line flux of $(1.75\pm0.16) \times 10^{-17}$ erg s$^{-1}$ cm$^{-2}$, a FWHM of $65.76\pm2.73$\AA\ (consistent with an unresolved line), and a line-flux signal-to-noise of 10.71 (see fitting results from Method 2 and Figure \ref{fig:Methods}). A summary of the properties of this emission line can be found in Table \ref{tab:1406}. 
The remaining lines which have been detected in only one of our two methods (Method 2) or both methods at a lower significance require further data to confirm their robustness, thus we are pursing ground-based spectroscopic follow-up to be discussed in a future paper.

\begin{figure*}[ht!]
{\includegraphics[width=\textwidth]{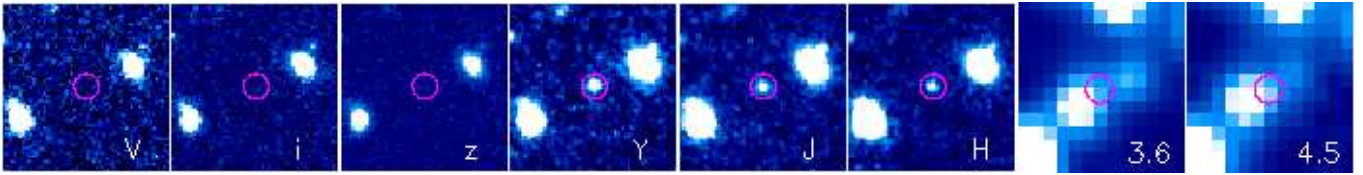}}
{\caption{Images of GS2\_1406 (circled in purple) from the CANDELS survey showing it to be a clear \textit{z}-band dropout. \textit{HST} images are 3.7"$\times$3.7" (61$\times$61 pixels), while \textit{Spitzer} images are 7.8"$\times$7.8" (13$\times$13 pixels). }\label{fig:Postage}}
\end{figure*}

\section{Discussion}
\subsection{Line Identification}
\label{sec:photometry}

%nJy photometry
%filters: f435w f606w f775w f814w f850l f098m f105w f125w f140w f160w irac3 irac4
%fluxes: -0.0000000      -7.7752427      0.48698310      -0.0000000       9.6749780       0.0000000       47.532145       48.569569       0.0000000       52.380569      -99.880805       229.38727
%errors: 1.0000000e+12       3.6331395       4.2779569   1.0000000e+12       5.2178261   1.0000000e+12       4.3888128       3.3656440   1.0000000e+12       4.1289827       70.255139       148.12831
%original: 12.982836       0.0000000       60.461557
%original errors: 5.2178261   1.0000000e+12       4.3888128
\begin{table*}[ht!]
\begin{tabular}{ |ccccccccc| }
 \hline
 \multicolumn{9}{|c|}{GS2\_1406 Photometric Measurements (in nJy)} \\
 \hline
 & V$_{606}$ & {\textit{i}}$_{775}$ & {\textit{z}}$_{850}$ & Y$_{105}$ & J$_{125}$ &  H$_{160}$ & 3.6$\mu$m & 4.5$\mu$m \\
 \hline
Measured Photometry &   $-7.78\pm3.6$ & $0.49\pm4.3$ & $12.98\pm5.2$ & $60.46\pm4.4$ & $48.57\pm3.4$ & $52.38\pm4.1$ & $35.44\pm49.4$ & $42.89\pm42.4$  \\
Line-Subtracted Values & - & - & $7.66\pm5.2$ & $39.48\pm4.4$ & - & - & - & - \\
\hline
\end{tabular}
\caption{Photometric measurements for GS2\_1406 in nJy, with the Ly$\alpha$ lineflux-subtracted values for the {\textit{z}}$_{850}$, Y$_{105}$ bands.  }\label{tab:phot}
\end{table*}

As our data set is derived from the high-redshift selected galaxies from the CANDELS-GOODS fields we have ample photometry measurements in these fields from \citet{Finkelstein2015}. Our emission-line galaxy, GS2\_1406 falls in the Hubble Ultra Deep Field (HUDF) second parallel field, referred to as the HUDF09-02 \citep{Bouwens2011b}. This field has deep WFC3 imaging from the HUDF09 survey (PI Illingworth; e.g. \citealt{Bouwens2010a, Oesch2010b}) and also has optical imaging with ACS \citep{Beckwith2006} from the UDF05 survey (PI: Stiavelli, \citealt{Oesch2007}).

This field has imaging in the V$_{606}$, {\textit{i}}$_{775}$, {\textit{z}}$_{850}$, Y$_{105}$, J$_{125}$, and H$_{160}$ bands, and was also observed with the \textit{Spitzer Space Telescope} Infrared Array Camera (IRAC; \citealt{Fazio2004}) program 70145 (the IRAC Ultra-Deep Field; \citealt{Labbe2013}) at 3.6$\mu$m and 4.5$\mu$m. Postage stamp images of this galaxy are shown in Figure \ref{fig:Postage} with the \textit{HST} images being 3.7"$\times$3.7" (61$\times$61 pixels), while \textit{Spitzer} images are 7.8"$\times$7.8" (13$\times$13 pixels). The galaxy is marked by a purple circle to show it being a clear \textit{z}-band dropout.

\begin{figure*}[ht!]
{\includegraphics{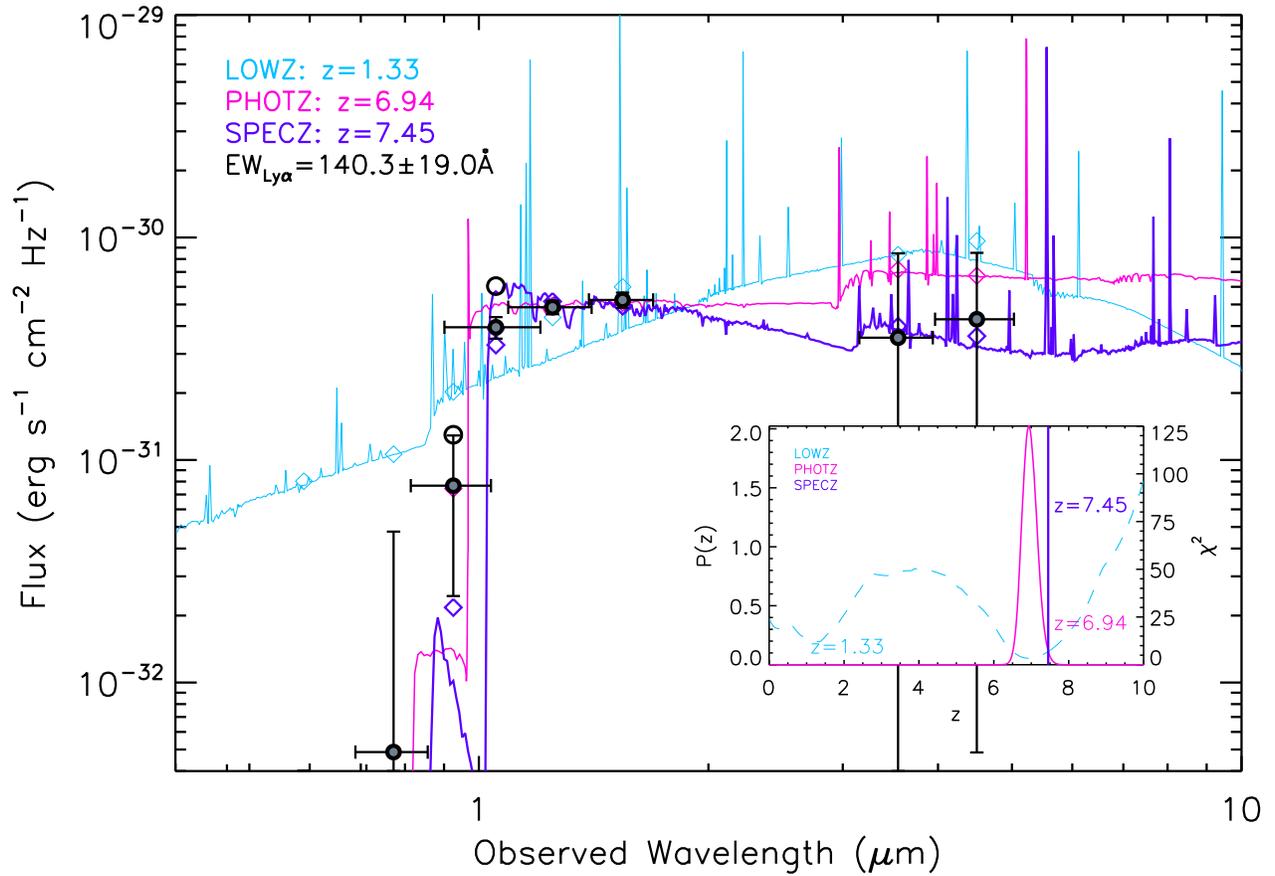}}
{\caption{Filled circles denote our measured photometry, removing the emission line contribution from the \textit{z}$_{850}$ and Y$_{105}$ bands (the flux values prior to this subtraction are shown by the open black circles). Horizontal error bars show the width of each filter through which 90\% of the flux is transmitted. The pink line shows the EAZY template at the best-fit photometric redshift of $z =$ 6.94, which is higher than the original CANDELS photo-z as we use the line-subtracted photometry values (we note this best fit template does include weak Ly$\alpha$ emission).  The purple and cyan lines show the best-fitting SED models at the spectroscopic redshift ($z =$ 7.452) and the potential lower-redshift photo-z solution ($z =$ 1.33), where these models come from full SED fitting the Ly$\alpha$-subtracted photometry (using models with no Ly$\alpha$ emission).  Colored diamonds are the corresponding SED flux for each filter. Inset: The PDF from our fiducial photometric redshift fit (pink), where the possible low-redshift solution is the second minimum in the chi-squared distribution, shown as the cyan dashed line.  The spectroscopic redshift is indicated by the purple vertical line.}\label{fig:SED}}
\end{figure*}

We used non PSF-matched catalogs for re-measuring the photometry values in elliptical Kron apertures, using the H$_{160}$ band as the detection image. We used an identical process as that done in \citet{Finkelstein2015}, measuring object colors in smaller apertures (PHOT\_AUTOPARAMS $=$ 1.2, 1.7), and then applying an aperture correction, based on the ratio between the default Kron aperture (PHOT\_AUTOPARAMS $=$ 2.5, 3.5) and that in our smaller aperture in the $H$-band. For the two bands impacted by our emission line at 1.03$\mu$m, the \textit{z}$_{850}$ and Y$_{105}$ bands, we subtract the contribution of the observed emission line from the measured photometry (see open circles in Figure \ref{fig:SED} for original photometry values).  While the aperture measurement in the \textit{z}-band shows a $\sim$ 2.5$\sigma$ significance measurement before subtraction of the Ly$\alpha$ lineflux, visual inspection of this region shows no significant connected pixels, implying that this measurement is likely dominated by random noise (as well as a $\sim$40\% flux contribution from our detected emission line). Prior to this subtraction, the $Y-J$ color from original photometry shows clear emission line contribution to the flux in that filter. 

\textit{Spitzer} IRAC photometry fluxes were originally deblended with T-PHOT \citep{Merlin2015}  measured by \citet{Finkelstein2015} and \citet{Song2016}, with a $> 2\sigma$ measurement in the 3.6$\mu$m band. However, this source is blended by two nearby sources.  To obtain the most robust measure of the flux at the position of our source in the IRAC images, we performed a dedicated deblended photometric measurement to the IRAC data by modeling the bright sources nearby using the GALFIT software \citep{Peng2002} and subtracting them from our image following a similar procedure as \citet{KFinkelstein2015}. We then use a 1.9'' circular aperture to measure the flux at the position of our object at the residual image. We use the photometric uncertainties from the T-PHOT catalog, as these accurately contain the uncertainty due to the residuals after subtracting the neighbor sources, and are conservatively larger than other uncertainty measures.  With these uncertainty values, we do not measure significant flux in the IRAC bands, consistent with visual inspection of the residual images. All photometric measurements for this galaxy can be found in Table \ref{tab:phot}.

%We initially make use of deblended TPHOT IRAC photometry from Song et al. (2016).  However, this source is blended by two nearby sources.  To obtain the most robust measure of the flux at the position of our source in the IRAC images, we ....stuff about galfit.  We use the photometric uncertainties from the TPHOT catalog, as these accurately contain the uncertainty due to the residuals after subtracting the neighbor, and are conservatively larger than other uncertainty measures.

\begin{table}[ht!]
\begin{tabular}{ |p{2.5cm}p{5.cm}| }
 \hline
 \multicolumn{2}{|c|}{GS2\_1406 Emission Line Values} \\
 \hline
 Coordinates		  &  (53.288090,-27.865408)  \\
 Peak Wavelength      &  $10280.60 \pm 3.94$\AA  \\
 Gaussian FWHM        &  65.76{\AA} $\pm$ 2.73{\AA}   \\
 Line Flux            &  $(1.75\pm0.16)\times10^{-17}$ erg s$^{-1}$ cm$^{-2}$  \\
 Signal-to-Noise      &  10.71 \\
 EW$_{Ly\alpha}$      &  140.3$\pm$19.0\AA \\
 \hline
\end{tabular}
\caption{Final emission line results for new z=7.542 Ly$\alpha$ detection in GS2\_1406 (CANDELS ID: z7\_PAR2\_2909). }\label{tab:1406}
\end{table}

We use these {\it HST}/ACS, {\it HST}/WFC3, and {\it Spitzer}/IRAC photometric measurements to measure the photometric redshift using the EAZY photometric redshift fitting code \citep{Brammer2008}.  EAZY measures a best-fit photometric redshift of $z =$ 6.94 with a secondary, low-redshift solution of $z =$ 1.33 obtained from the second $\chi^{2}$ minimum (see inset of Figure \ref{fig:SED}). Spectral energy distributions (SEDs) of best fit galaxy templates at these redshifts are plotted in Figure \ref{fig:SED}. The fiducial photometric redshift in pink is EAZY's best fit solution, while the spectroscopic redshift (purple) and low-redshift solution (blue) are best-fit templates at those redshifts. If the low-redshift solution were correct, the observed  emission line could be instead [O\,{\sc ii}] at $z =$ 1.30.  However, as shown in Figure \ref{fig:SED}, a galaxy at this redshift would be expected to have emission significantly higher than the observed limits in the optical bands with no significant spectral break, and a much redder SED in the detected bands, thus we consider the low-redshift solution to be ruled out.

Our interpretation is thus that the detected line is Ly$\alpha$ at a spectroscopic redshift of $z = 7.452\pm0.003$.  This deviates from the photometric redshift best-fit solution of $z =$ 6.94 at the $\sim 2\sigma$ level.  To see if this difference is due to the EAZY template set and/or fitting method, we verify this result using the Bayesian Photo-Z estimation (BPZ; \citealt{Benetiz2000}), and get a similar photometric redshift measure of $z =$ 6.8. This discrepancy is not necessarily a problem, as photometric redshifts have not been spectroscopically calibrated at high redshifts, and even so, 2$\sigma$ deviations are expected $\sim$5\% of the time. Clearly a larger number of spectroscopic redshifts are needed to validate photometric redshift probability density functions (PDFs) as photometric redshifts are fundamentally not right for single objects since they primarily rely on templates. As no two galaxies are truly identical, it is not surprising that photometric redshifts work well for looking at properties of many galaxies but can fail in individual cases. If significant outliers like these are found to be commonplace, it would imply that our photometric redshift uncertainties are higher than expected, resulting in increased uncertainties in luminosity functions and galactic properties. Discussion of this is covered in more detail in \citet{Pirzkal2017}.

To derive relevant galaxy physical properties, we performed galaxy SED fitting with the line-subtracted HST/ACS and WFC3 and Spitzer/IRAC fluxes and thus, the models do not have Ly$\alpha$ emission. Our SED fitting is based on a MCMC algorithm and uses the \citet{Bruzual2003} stellar population synthesis model, and the details of the SED fitting is described in \citet{Jung2017}. We find that the 68\% confidence measurements of this object give a stellar mass of log(M/M$_{\odot}$) = 8.79 to 8.99, and a dust-corrected UV star-formation rate of 7.77 to 8.32 M$_{\odot}$ yr$^{-1}$.  The SED is also fairly blue, thus the model fitting unsurprisingly prefers little dust with an (E[B-V] = 0.007 to 0.057).

%From Intae's SED Fitting
%68%-confident ranges are as below.
%========================================
%Log(M/M_sun) :    8.5725534 -        8.9917991
%Log(Age/yr)   :       7.7789697 -        8.3122996
%E(B-V)       :    0.0070697928 -      0.056505023
%SFR(M_sun/yr):          5.1558547 -        8.4519820
%continuum @ 1216-1316A [erg s^-1 cm^-2 A^-1] :    0.12471738
%M1500             :      -19.931802

% \begin{table}[ht!]
% \begin{tabular}{ |p{2.5cm}p{5.cm}| }
%  \hline
%  \multicolumn{2}{|c|}{GS2\_1406 Derived Properties} \\
%  \hline
%  M$_{\rm AB}$		  &  $XX \pm XX$ \\
%  SFR$_{\rm UV}$       &  $XX \pm XX$ M$_{\odot}$ yr$^{-1}$  \\
%  %SFR$_{\rm Ly\alpha}$ &  $XX \pm XX$ M$_{\odot}$ yr$^{-1}$  \\
%  $\beta_{\rm UV}$     &  $XX \pm XX$ \\
%  R$_{\rm gal}$            &  $XX \pm XX$ kpc \\
%  \hline
% \end{tabular}
% \caption{Derived galaxy properties for GS2\_1406 (CANDELS ID: z7\_PAR2\_2909). }\label{tab:prop}
% \end{table}

\subsection{Ly$\alpha$ Equivalent Width}
\label{sec:ew}

 \begin{figure}[ht!]
 {\includegraphics[width=8.5cm]{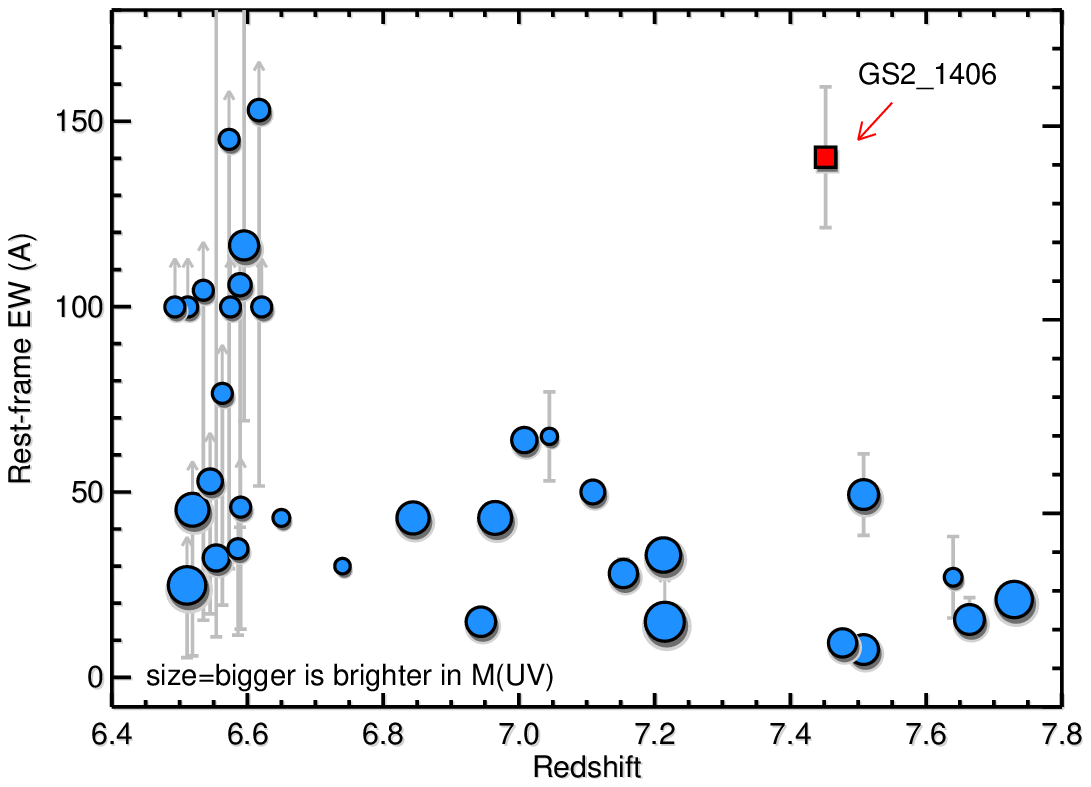}}
 {\caption{This shows the redshift evolution of rest frame Ly$\alpha$ equivalent width (EW$_{Ly\alpha}$) for galaxies with high spectroscopic confidence (\citealt{Iye2006, Ouchi2010, Schenker2012, Vanzella2011, Ono2012, Rhoads2012, Finkelstein2013, Pentericci2014}. There was a missing population of high EW$_{Ly\alpha}$ galaxies at $z >$ 7 prior to the detection of this GS2\_1406 (red square) which falls squarely in the high EW$_{Ly\alpha}$, high redshift range with an EW$_{Ly\alpha}$ = 140.3$\pm$19.0\AA. Adapted from \citet{Tilvi2014}  }\label{fig:zvM}}
 \end{figure}
 
Ly$\alpha$ rest-frame equivalent width (EW$_{Ly\alpha}$) measurements are much lower at $z >$ 7, possibly due to an increase in the neutral fraction of the IGM \citep{Forero-Romero2012, Tilvi2014}. However, recent observations by \citet{Hu2017} and \citet{Zheng2017} have found luminous Ly$\alpha$ emitters at $z>7$. The EW$_{Ly\alpha}$ measurement for GS2\_1406 is taken by comparing the grism-measured line flux to the best fit SED template at the spectroscopic redshift (z=7.452) immediately red-ward of the Ly$\alpha$ line (average of 1220-1320 \AA\ rest-frame) as the continuum value. The continuum flux is $1.25\times 10^{-19}$ erg s$^{-1}$ cm$^{-2}$ \AA$^{-1}$, giving GS2\_1406 a EW$_{Ly\alpha}$ = 140.3$\pm$19.0\AA, much higher than any previously spectroscopically confirmed galaxy at $z >$ 7, as shown in Figure \ref{fig:zvM}. While robust statements about the ionization state of the IGM cannot be made from one galaxy, either internal galaxy kinematics or ionized bubbles must conspire to allow for the presence such bright Ly$\alpha$ in this galaxy.

%Galaxy  LineFlux dLineFlux Lambda dLambda z dz  Sig dSig  snr  chisq
 % GS2_1406    1.75186145e-17    1.61704321e-18    10280.59960938        3.94470119        7.45444059        0.00324345       27.94210052        1.27992237

\section{Summary}
\label{sec:summary}
Using the deepest {\it HST} Grism data available we have built an automated detection method to find emission lines from CANDELS-GOODS continuum-selected $z >$ 5.5 galaxies. This data includes 5 separate roll angles to reduce the impact of contamination, and we then perform additional reduction to remove any residual contamination in our spectra. We searched for $> 4\sigma$ emission lines using two different methods. In the first method, we compare the results for each galaxy across all roll angles and identify significant lines as those which are detected at the same wavelength in more than one roll angle. This method finds 2 emission-line galaxies. In the second method, we perform a fit to all five roll angles simultaneously, using a combined $\chi^{2}$ value, which finds 5 emission-line galaxies. Of these two, one is a previously measured Ly$\alpha$ line \citep{Finkelstein2013}, already extensively studied in this data set by \citet{Tilvi2016}, and our routine recovers the same line flux as previously reported. The other is a first-time detection in GS2\_1406, discovered photometrically as z7\_PAR2\_2909 by \citet{Finkelstein2015}. 

GS2\_1406 has a detected emission line at $\sim$1.03$\mu$m, a line flux of $(1.75\pm0.16) \times 10^{-17}$ erg s$^{-1}$ cm$^{-2}$, and a line-flux signal-to-noise of 10.71. We compare this result with broadband photometric measurements of this galaxy, and interpret this line to be Ly$\alpha$ at a redshift of $z= $ 7.452.  This spectroscopic redshift is a 2$\sigma$ outlier from the photometric redshift ($z =$ 6.94) illustrating the caveats of simple photometric redshift determinations for single sources. If further follow-up on other emission line galaxies in this data set show a similar offset, this could have strong implications on the accuracy of photometric redshift fitting.

This galaxy also has the highest Ly$\alpha$ rest-frame equivalent width (EW$_{Ly\alpha}$) at $z >$ 7: 140.3$\pm$19.0\AA. It is expected that EW$_{Ly\alpha}$ should decrease with $z$, paralleling an increase in the neutral fraction of the IGM during the epoch of reionization. The consequence of finding a high-redshift, high-EW$_{Ly\alpha}$ galaxy could mean there is a highly ionized line-of-sight to this galaxy, or that the kinematics in this galaxy result in Ly$\alpha$ being emitted significantly red-ward of the systemic redshift.  These scenarios, as well as a higher confidence in the line identification, can be obtained with higher-resolution follow-up of the Ly$\alpha$ line and measurement of another emission line such as rest-UV C\,{\sc iii}] or rest-FIR [C\,{\sc ii}]. \\

\acknowledgments
Based on observations made with the NASA/ESA \textit{Hubble Space Telescope}, obtained [from the Data Archive] at the Space Telescope Science Institute, which is operated by the Association of Universities for Research in Astronomy, Inc., under NASA contract NAS 5-26555. These observations are associated with program \#13779. RL and SF acknowledge support provided by NASA through a grant from the Space Telescope Science Institute, which is operated by the Association of Universities for Research in Astronomy, Inc., under NASA contract NAS 5-26555.  RL also acknowledges support from the National Science Foundation through the MPS-GRSV program under grant number 1707552. AC acknowledges the grants ASI n.I/023/12/0 "Attivit\'a relative alla fase B2/C per la missione Euclid" and PRIN MIUR 2015 "Cosmology and Fundamental Physics: Illuminating the Dark Universe with Euclid". LC is supported by grant ID DFF-4090-00079. This work was supported by  grants HST GO-13779.* from STScI, which is operated by AURA for NASA under contract NAS 5-26555. RAW acknowledges support from NASA JWST Interdisciplinary Scientist grants NAG5-12460 and NNX14AN10G from GSFC.

%\facility{facility ID}
\facilities{\textit{HST}, \textit{Spitzer}, TACC} 
\software{IDL, Python, EAZY, BPZ, SourceExtractor}

\bibliographystyle{yahapj}
\bibliography{references}

%\appendix
%\section{appendix section}

\end{document}